\documentclass{article}
\usepackage{graphicx}
\begin{document}
\title{Dynamic Scaling of an Adsorption-Diffusion Process on Fractals}
\author{Hidetsugu Sakaguchi,\\
Department of Applied Science for Electronics and Materials, \\
Interdisciplinary Graduate School of Engineering Sciences, \\
Kyushu University, Kasuga, Fukuoka 816-8580}
\maketitle

\begin{abstract}
  We study numerically a dynamic scaling of a diffusion process involving the Langmuir type adsorption. 
We find dynamic scaling functions in one and two dimensions and compare them with direct numerical simulations. We study further the dynamic scaling law on fractal surfaces. The adsorption-diffusion process obeys the fracton dynamics on the fractal surfaces.
\end{abstract}    
\bigskip
\section{Introduction}
Diffusion and adsorption are fundamental processes in various nonequilibrium phenomena, especially in the research fields of fractals~\cite{rf:1}. 
The diffusion-limited-aggregation (DLA) is a typical fractal pattern generated by the diffusion and deposition \cite{rf:2,rf:3}.  The fractal growth and anomalous scaling have been studied even in atomistic processes and nanoscale depositions of thin films \cite{rf:4,rf:5,rf:6}.  Transport phenomena on various fractals were also intensively studied, and the anomalous dynamical behavior is characterized by the fracton dimension \cite{rf:7,rf:8}. Irreversible reaction-diffusion systems such as A$+$A$\rightarrow$ O or A$+$B$\rightarrow$ O were also studied on fractal substrates \cite{rf:9,rf:10}. 

Adsorption phenomena on fractal surfaces are important in chemistry \cite{rf:11}.  The adsorption on porous media such as activated carbons, zeolites and silica-gels is  used for elimination, separation, and storage of various materials. 
The fractal dimensions of such porous media were measured by the adsorption of particles with various sizes, the small-angle x-ray scattering and the scanning tunneling microscopy \cite{rf:12,rf:13}. The transport phenomena including the diffusion process are important for the temporal evolution of the adsorption onto the porous materials. 
The diffusion and adsorption processes on fractals are important also in biology \cite{rf:14}. The oxygen is transported to the whole body through ramifying structures of lungs and blood vessels and then adsorption and chemical reactions occur at the ends. The fractal dimensions were estimated for lungs and blood vessels \cite{rf:1,rf:14,rf:15}. The convective transport by fluid flow plays an important role on a larger scale, but the diffusion processes are important in smaller scales, because the fluid velocity decays by viscosity. 

In this paper, we study a simple adsorption and diffusion process on regular lattices and typical fractals.  We consider a reversible adsorption process and the total mass is locally conserved in the time evolution. Our model is a different type problem from irrevesible adsorption or the reaction-diffusion models studied  before \cite{rf:9,rf:10,rf:16}. 
We use a deterministic model and do not consider statistical fluctuations. We obtain dynamic scaling functions with  numerical simulations. 
Although the model and the simulations are very simple, such scaling functions for the adsorption process were not reported before.

\section{Model equation and dynamic scaling on regular lattices}
Firstly, we consider a one-dimensional adsorption-diffusion model. 
We consider the Langmuir type adsorption \cite{rf:17}. The rate of adsorption is proportional to the density $c$ of the diffusion particle and the number of vacant sites for adsorption. The occupation fraction $\theta$ ($0<\theta<1$) of the surface site obeys 
\begin{equation}   
\frac{d\theta}{dt}=k_ac(1-\theta)-k_d\theta,
\end{equation}
where $k_a$ and $k_d$ are the rate of adsorption and desorption respectively.  
At equilibrium, the fraction $\theta$ is given by $\theta_0=bc/(bc+1)$, where $b=k_a/k_d$.  
The diffusion particle obeys a reaction-diffusion equation
\begin{equation}
\frac{\partial c}{\partial t}=-k_ac(1-\theta)+k_d\theta+D\frac{\partial^2c}{\partial x^2},\end{equation}
where $D$ is the diffusion coefficient. The conservation law of $c+\theta$ is locally satisfied in Eqs.~(1) and (2). 
We study a simple adsorption-diffusion process where the particle density $c$ is fixed to a constant value $c_0$ at $x=0$, that is, the system is in contact with a particle reservoir. The initial conditions are assumed to be $c(x,0)=0$ and $\theta(x,0)=0$, that is, there are no adsorbates and diffusion particles  initially. 
We have performed a direct numerical simulation of eqs.~(1) and (2) for $c_0=1, k_a=1, k_d=0.5, D=1$ and $L=1000$. Figure 1(a) displays three snapshots of $c(x,t)$ at $t=50, 200$ and 800. The particles penetrate into the adsorbent as time goes. Figure (1b) displays five snapshot profiles of $c(x,t)$ at $t=200,400,600,800$ and 1000 by rescaling the horizontal axis by $x^{\prime}=x/t^{1/2}$.  The perfect overlapping implies that the time evolution of $c(x,t)$ obeys a scaling law $c(x,t)=c(x/t^{1/2})$ and $\theta(x,t)=\theta(x/t^{1/2})$. If the scaling law is assumed, eqs.~(1) and (2) are rewritten as 
\begin{equation}
-1/2x^{\prime}\frac{dc}{dx^{\prime}}=D\frac{d^2c}{dx^{\prime 2}}+1/2x^{\prime}\frac{d\theta}{dx^{\prime}},
\end{equation}
where $x^{\prime}=x/t^{1/2}$.  Substitution of the equilibrium condition $\theta=bc/(1+bc)$ into eq.~(3) yields
\begin{equation}
D\frac{d^2c}{dx^{\prime 2} }=-1/2x^{\prime}\left (1+\frac{b}{(1+bc)^2}\right )\frac{dc}{dx^{\prime}}.
\end{equation}
\begin{figure}[tbh]
\begin{center}
\includegraphics[height=3.5cm]{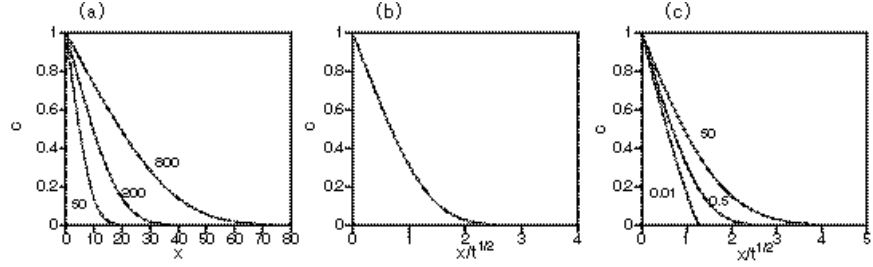}
\end{center}
\caption{(a) Three snapshot profile of $c(x,t)$ at $t=50,200$ and 800. (b) Five profiles of $c(x/t^{1/2})$ at $t=200n$ ($n=1,\cdots,5$) (Solid curves), and a solution of eq.~(2) (dashed line). They are completely overlapped. (c) Three dynamic scaling functions for $b=0.01,0.5$ and 50.} \label{fig1}
\end{figure}
\begin{figure}[tbh]
\begin{center}
\includegraphics[height=3.5cm]{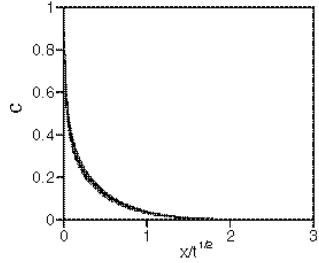}
\end{center}
\caption{Seven profiles of $c(x/t^{1/2})$ at $t=10+50n$ ($n=1,\cdots,7$) (Solid  curves), and a solution of eq.~(3) (dashed line). They are well overlapped.} \label{fig2}
\end{figure}
A solution to the nonlinear differential equation (4) approaches 0 for $x^{\prime}\rightarrow \infty$, only if the derivative $\partial c/\partial x^{\prime}$ takes a special value at $x^{\prime}=0$. The solution can be numerically obtained by searching the special value. The dashed curve in Fig.~(1b) is the special solution, and the solution represents the scaling function $c(x/t^{1/2})$.  If $k_a=k_d=0$, the adsorption and desorption do not occur, then the scaling function is exactly solved as 
\begin{equation}c(x^{\prime})=\frac{c_0}{\sqrt{\pi D}}\int_{x^{\prime}}^{\infty}\exp(-x^2/(4D))dx.
\end{equation}
The numerically obtained scaling function for $k_a=1$ and $k_d=0.5$ is definitely smaller than the scaling function (5), because the adsorption process decreases the density of the diffusion particle. Figure 1(c) displays the dynamic scaling functions at $b=0.01,0.5$ and 50. As $b$ is increased, the dynamic scaling function has a longer tail. The dynamic scaling function at $b=50$ is very close to the function of eq.~(5).   

We have found a similar dynamic scaling law in the two-dimensional system.  The particle density is fixed to be constant $c_0$ at the central point $(x,y)=(L/2,L/2)$. We have performed numerical simulations on a square lattice for $k_a=1,k_d=0.5,D=1$ and $c_0=1$, and the system size is $L\times L=600\times 600$. In the two dimensional system, the last diffusion term in eq.~(2) is replaced by $c(i+1,j)+c(i-1,j)+c(i,j+1)+c(i,j-1)-4c(i,j)$, where $(i,j)$ denotes a lattice site in the square lattice. Figure 2 displays seven snapshot profiles of $c(x^{\prime})$ at  $t=10+50n$ ($n=1,\cdots,7$), where $x^{\prime}=(i-L/2)/t^{1/2}$ and $j$ is fixed to be $L/2$. 
The dynamic scaling law $c(i,j,t)=c(r/t^{1/2})$ is satisfied for large $t$, where $r=\sqrt{(i-L/2)^2+(j-L/2)^2}$ is the distance from the center. The two-dimensional scaling function is expected to obey            
\begin{equation}
D\frac{d^2c}{dx^{\prime 2}}=-\left (D/x^{\prime}\right)\frac{dc}{dx^{\prime}}-1/2x^{\prime}\left (1+\frac{b}{(1+bc)^2}\right )\frac{dc}{dx^{\prime}},
\end{equation}
where $x^{\prime}=r/t^{1/2}$, if the two-dimensional Laplacian is replaced by $\partial^2/\partial r^2+(1/r)\partial/\partial r$.  A numerically obtained solution which approaches to 0 for $r\rightarrow \infty$ is displayed by a dashed line in Fig.~2. 
The derivative $dc/dx^{\prime}$ diverges as $1/x^{\prime}$ at $x^{\prime}=0$.
The theoretically estimated dynamic scaling function is a good approximation to the scaling function obtained by direct numerical simulations. 

\section{Dynamic scaling of a diffusion-adsorption process on fractal substrates}
\begin{figure}[tbh]
\begin{center}
\includegraphics[height=4.cm]{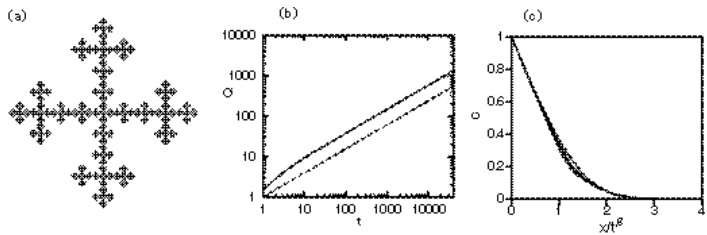}
\end{center}
\caption{(a) Vicsek's fractals with fractal dimension $\log5/\log 3$. (b) Time evolution of the total adsorbent quantity $Q(t)$. The dashed line denotes the curve with exponent $\alpha=\log 5/(\log 3+\log 5)$ (c) Dynamic scaling of $c(x/t^{\beta})$ (solid lines) where $\beta=1/d_w=\log 3/(\log 3+\log 5)$. The dashed line is an approximate function using eq.~(7).} \label{fig3}
\end{figure}
We investigate further dynamic scaling laws in the adsorption-diffusion process on fractal substrates. The first example is a deterministic fractal set of fractal dimension $D_f=\log 5/\log 3\sim 1.46$ as shown in Fig.~3(a), which is called Vicsek's fractals \cite{rf:18}.  The size of the fractal set we used is $L=3^6=729$, that is, the contraction map is repeated 6 times. The diffusion process occurs only on the fractal set. 
Namely, the diffusion term $\sum_{i^{\prime},j^{\prime}}\{c(i^{\prime},j^{\prime})-c(i,j)\}$ is calculated only for the nearest-neighbor sites on the fractal set.
Figure 3(b) displays the time evolution of the total quantity $Q$ of adsorbed particles $\sum_{i,j}\theta_{i,j}(t)$. The quantity grows as $t^{\alpha}$ where $\alpha$ is close to $\log 5/(\log 5+\log 3)$. The exponent is evaluated from the relation $Q\sim R^{D_f}\sim (t^{1/d_w})^{D_f}=t^{D_f/d_w}$, where $R$ is the typical radius of the extent of the diffusion particle and $d_w$ is the anomalous-diffusion exponent, and the relation $R^2\sim t^{2/d_w}$ is used. The exponent $d_w$ is 2 in one and two dimensions. The exponent $\alpha$ is equal to $D_s/2$, where $D_s=2D_f/d_w$ is the fracton dimension. For the loopless fractal, $d_w$ is given by $D_f+1$ \cite{rf:7,rf:19}, therefore $\alpha$ is evaluated as $\alpha=D_f/(D_f+1)=\log 5/(\log 5+\log 3)$. That is, the dynamic exponent $\alpha$ is determined by the anomalous diffusion on the fractal set. This is due to the conservation law for the diffusion particles and the adsorbed particles. However, the form of the dynamic scaling function depends on the detail of the adsorption process. Figure 3(c) displays four snapshots of $c(x/t^{1/d_w})$ at $t=20000+500n$ ($n=1,\cdots,4$), which shows the dynamic scaling with exponent $\beta=1/d_w=\log 3/(\log3+\log 5)$. If the scaling function of the form $c(x/t^{1/d_w})$ is assumed, and the diffusion constant $D$ is assumed to behave effectively as $D(t)\sim D_0t^{2\beta-1}$ \cite{rf:7,rf:11}, eq.~(4) is rewritten as   
\begin{equation}
D_0\frac{d^2c}{dx^{\prime 2}}=-(1/d_w)x^{\prime}\left (1+\frac{b}{(1+bc)^2}\right )\frac{dc}{dx^{\prime}}.
\end{equation}
The dashed line is a solution of eq.~(7) for $1/d_w=\log3/(\log 3+\log 5)$ and $D_0=1$. 
It is close to the dynamic scaling function obtained by direct numerical simulations, but small deviation is observed, probably because the diffusion term on the fractal set is approximated by the second derivative $D(t) d^2c/dx^{\prime 2}$ with the time dependent diffusion constant. 
\begin{figure}[tbh]
\begin{center}
\includegraphics[height=3.5cm]{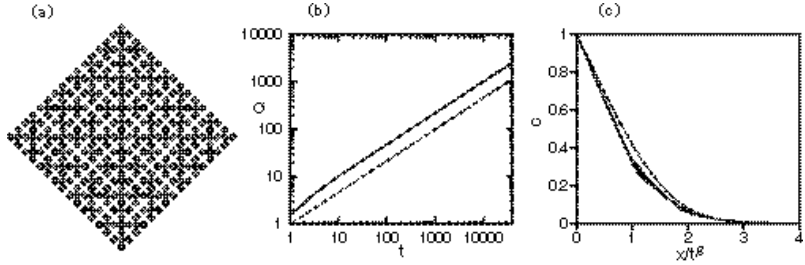}
\end{center}
\caption{(a) Loopless fractal with fractal dimension 2. (b) Time evolution of the total adsorbent quantity $Q(t)$. The dashed line is a curve with exponent of $\alpha=2/3$. (c) Dynamic scaling of $c(x/t^{\beta})$ (solid lines) where $\beta=1/d_w=1/3$. The dashed line is an approximate function using Eq.~(7).} \label{fig4}
\end{figure}

The second example is another deterministic loopless fractal of dimension 2 as shown in Fig.~4(a). The dimension is exactly 2, but there are no loops, in contrast to the normal square lattice. Therefore, the anomalous diffusion exponent is $d_w=3$, which is definitely different from 2 for the normal two-dimensional system, and the fracton dimension is 4/3. 
Figure 4(b) displays a time evolution of the total quantity $Q(t)$ of the adsorbed particles, which grows approximately $Q(t)\sim t^{2/3}$.       
Figure 4(c) displays four snapshots of $c(x/t^{1/d_w})$ at $t=20000+500n$ ($n=1,\cdots,4$), which shows the dynamic scaling function with exponent $1/d_w=1/3$.  The dashed line is a solution which satisfies eq.~(7) for $1/d_w=1/3$ and $D_0=1$. 
There is some difference between the solution to eq.~(7) and the numerically obtained  scaling function in this case. 

\begin{figure}[tbh]
\begin{center}
\includegraphics[height=3.5cm]{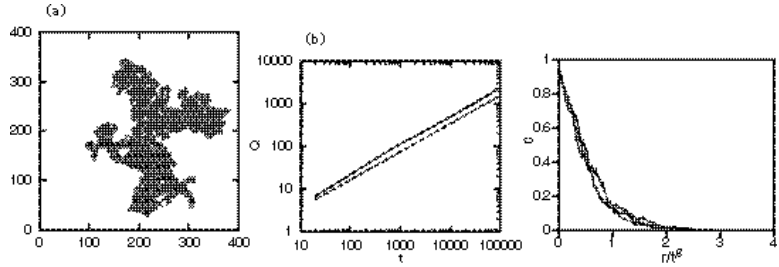}
\end{center}
\caption{(a) Adsorbed sites satisfying $\theta>0.01$ at $t=100000$. (b) Time evolution of the total quantity $Q(t)$ of the adsorbed particles. The dashed line is a curve with exponent of $\alpha=0.66$. (c) Dynamic scaling of $c(r/t^{\beta})$ at $t=12000,40000$ and 100000, where $\beta=1/2.87$} \label{fig5}
\end{figure}
The third example is an adsorption-diffusion process on a critical percolation cluster. 
The critical site percolation cluster is constructed on a square lattice. The diffusion and the adsorption occur only on the percolation cluster. The density of $c$ is fixed to be 1 at a certain site $(200,201)$. Figure 5(a) displays the sites where the adsorbate quantity $\theta$ is larger than 0.01 at $t=100000$. Figure 5(b) displays a time evolution of the total quantity $Q(t)$ of the adsorbed particles, which is approximated by $Q(t)\sim t^{0.66}$. The fractal dimension of the two dimensional percolation cluster is $D_f=91/48$ and the anomalous diffusion exponent $d_w$ is known to be $d_w\sim 2.87\pm 0.02$ \cite{rf:20}.   The fracton dimension $d_s=2D_f/d_w\sim 1.31\pm 0.02$, which is close to the Alexander and Orbach conjecture $d_s=4/3$ for the percolation clusters in any dimensions \cite{rf:21}. Our adsorption-diffusion process obeys the dynamic scaling law with the same exponent as the fracton dynamics on the percolation cluster.  Figure 5(c) displays three profiles of $c(r/t^{1/2.87})$ at $t=12000,40000$ and 100000, where $r$ is the distance from the center $(200,201)$ and the average of $c$ is taken for all the sites on the percolation cluster, which are located between two circles of the radius $r$ and $r+1$. The three plots are overlapping fairly well, which implies that the dynamic scaling law is satisfied.

\begin{figure}[tbh]
\begin{center}
\includegraphics[height=4.cm]{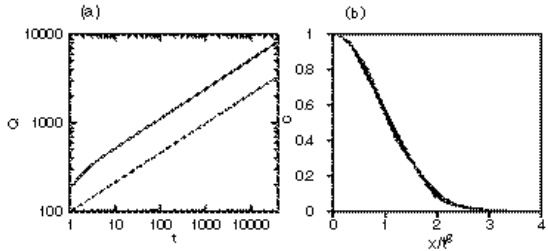}
\end{center}
\caption{(a) Time evolution of the total quantity $Q(t)$ of the adsorbed particles for the loopless fractal shown in Fig.~4(a). On the lower-left side, the density $c$ is fixed to be 1. The dashed line is a curve with exponent of $\alpha=1/3$. (c) Dynamic scaling of $c(x/t^{\beta})$ at $t=20000,25000,30000,35000$ and 40000, where $\beta=1/3$} \label{fig6}
\end{figure}
We have performed several simulations in which the particle density is fixed at a point for the sake of simplicity.
However, adsorptive particles usually penetrate from the surface of the porous media. We have performed a few numerical simulations imposing the fixed value $c_0$ on a line. As an example, we have used a fractal substrate shown in Fig.~4(a), and the particle density is fixed to be $c(x,y)=c_0$ at the lower-left side satisfying $x+y=257$. On the other three boundary sides, no-flux boundary conditions are imposed. We have calculated the total quantity $Q$ of the adsorbed particles and the time evolution of $c(x,y)$ on the line satisfying $y=256$. Figure 6(a) displays a time evolution of $Q$ (solid line), which obeys $Q(t) \sim t^{1/3}$. The exponent $\alpha$ is reduced by half in contrast to the value shown in Fig.~4(b), since the adsorption-diffusion process evolves in a quasi-one-dimensional manner.
Figure 6(b) displays the dynamic scaling of $c(x/t^{\beta})$ with the exponent $\beta=1/d_w=1/3$, which is the same exponent as the one in Fig.4~(c). 
The exponent $\beta$ does not depend on the boundary condition.  

\begin{figure}[tbh]
\begin{center}
\includegraphics[height=3.5cm]{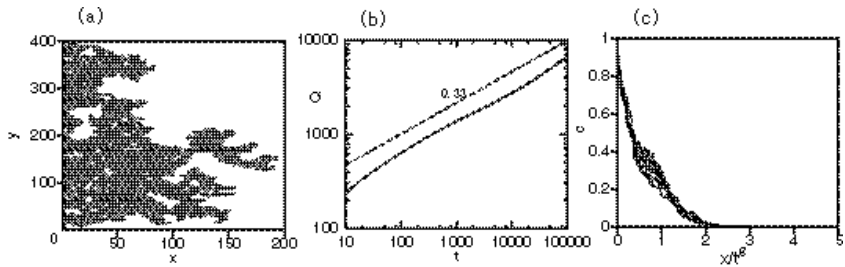}
\end{center}
\caption{(a) A snapshot pattern of adsorbed sites satisfying $\theta>0.01$ at $t=100000$. On the left side $x=0$, the density $c$ is fixed to be 1, in contrast to the simulation in Fig.~5. (b) Time evolution of the total quantity $Q(t)$ of the adsorbed particles on the percolation cluster.  The dashed line denotes  a line with exponent of 0.33. (c) Dynamic scaling of $c(x/t^{\beta})$ at $t=20000\cdot n$ ($n=1,\cdots, 5$) with $\beta=1/2.87$} \label{fig7}
\end{figure}
We have performed a similar simulation on the random lattice contructed by the site percolation used in Fig.~5. The particle density is fixed at the left side $x=0$ and the no-flux boundary conditions are imposed on the other three sides.   Figure 7(a) displays the region with $\theta>0.01$ at $t=100000$, which shows the adsorptive particles penetrate into the percolation cluster from the left side. Figure 7(b) displays the total quantity $Q(t)$ of the adsorbed particles. $Q(t)$ roughly obeys a power law $Q(t)\sim t^{\alpha}$ with $\alpha=0.33$. The exponent $\alpha$ is reduced by half, compared to the value shown in Fig.~5(b) owing to a quasi-one-dimensional growth from the linear surface.   Figure 7(c) displays five profiles of $c(x/t^{1/2.87})$ at $t=20000\cdot n$ ($n=1,2,\cdots,5$), where the average of $c$ is taken over the $y$ coordinate on the percolation cluster. The five plots are overlapping fairly well, which implies that the dynamic scaling law with the $\beta=1/2.87$ is satisfied. The exponent is the same as the one in Fig.~5(c). Thus, we have found that the exponent $\alpha$ of the total adsorption $Q(t)$ depends on the boundary conditions, but the exponent $\beta$ for the dynamic scaling function does not depend on the boundary conditions.    

\section{Summary}
We have performed direct numerical simulations of a simple and instructive model equation for the diffusion and adsorption process on regular lattices and regular two-dimensional fractals, and a two-dimensional percolation cluster.  We have found the dynamic scaling laws in two kinds of simulations with different fixed boundary conditions.  
The total quantity of the adsorbed particles obeys a power law $Q(t)\sim t^{\alpha}$, where the exponent is $d/2$ for one and two dimensional systems and $\alpha\sim D_f/d_w$ on fractal sets, when the particle density is fixed at a certain point.  The number of surface sites (the surface area) is characterized by the fractal dimension $D_f$. It is favorable for the adsorption that the surface area is enlarged owing to the fractal surface. 
However, the diffusion occurs more slowly on fractal surfaces than on the smooth surfaces, which is unfavorable for the fast adsorption. It is characterized by the anomalous dynamic exponent $1/d_w$. The exponent $\alpha$, which determines the time evolution of the total quantity of the adsorption, is given by the product of  $D_f$ and $1/d_w$, when the particle density is fixed at a certain point. The exponent $\alpha$ is halved when the particle density is fixed at a certain line.  
The dynamic scaling functions $c(x/t^{1/d_w})$ were found on one and two dimensional lattices and several fractals. We have found the form of the dynamic scaling function for one and two dimensional lattices by solving numerically Eqs.~(4) and (5). 

We have tried to get an approximate dynamic scaling function for the adsorption-diffusion process on fractals, however, 
it is left to future to obtain a better form of the dynamic scaling function for the adsorption-diffusion processes on fractals.
Our model is a very simple deterministic model which corresponds to the mean-field approximation, however, the applicability to real systems might be limited, since statistical fluctuations, interactions between adsorbed particles and convective flows are neglected.  It is desirable to generalized the simplest model to include such various effects. 
It is also left to study the diffusion-adsorption on realistic three-dimensional random fractals using other type of adsorption process such as the BET type adsorption \cite{rf:22}.

\end{document}